\documentclass{ws-p9-75x6-50}

\def\Journal#1#2#3#4{{#1} {\bf #2}, #3 (#4)}

\def\RevRou{\em Rev. Roum. Sci. Tech.- M\'ec. Appl.}
\def\Arch{\em Arch. Mech.}
\def\ZfN{\em Z. Naturforsch.}
\def\AnBuc{\em Analele Univ. Bucuresti- Fizica}

\newcommand{\drond}[2]
      {\frac{\textstyle{\strut \partial#1}}{\textstyle{\strut \partial #2}}}

\begin{document}

\title{ON A SCALAR THEORY OF GRAVITATION}

\author{Mayeul ARMINJON}

\address{Laboratoire "Sols, Solides, Structures" [U.M.R. 5521 du CNRS]\\BP 53, 38041 Grenoble cedex 9, France}

\maketitle

\noindent Our motivation was to extend the Lorentz-Poincar\'e
ether theory so that it could describe gravitation. The latter
theory consists in assuming that Maxwell's equations are valid in
some fundamental inertial frame or "ether," and that each material
object that moves through the ether undergoes a Lorentz
contraction. This is physically equivalent to Einstein's special
relativity (SR),~\cite{prokhovnik93} but it differs from standard
SR at the meta-physical (interpretation)
level.~\cite{prokhovnik93} The construction of the scalar theory is based
on a tentative concept of physical vacuum as a space-filling
perfect fluid, or "micro-ether." This concept leads to a definite
set of equations: it is this set that should be assessed from the
predictions it leads to, and from the comparison of these
predictions with experimental data. According to that concept,
material particles would be organized flows in that fluid, such as
vortices, thus each particle would occupy some bounded domain in
the fluid. The gravitational force is interpreted as resulting
from the forces exerted on any such "particle" (domain), due to
the spatial variation of the fluid pressure over macroscopic
distances. This leads to define a gravity acceleration vector as
follows:~\cite{arminjon93a}
\begin{equation}
{\sf g}= -\frac{grad\: p_e}{\rho_e}, \label{eq:vecteur-g}
\end{equation}
where $p_e$ and $\rho_e = \rho_e(p_e)$ are the macroscopic fields of pressure and density in
the imagined fluid (micro-ether). Note that this equation implies that $p_e$ and $\rho_e$ {\em
decrease} towards the gravitational attraction. The {\em preferred reference frame} of the
theory is that one which is obtained by averaging the velocity field of the micro-ether over a
very large scale. An equation for the scalar gravitational field $p_e$ (or equivalently the
field $\rho_e$) follows from the requirement that Newtonian gravity should be recovered if the
micro-ether were an incompressible fluid, and from the analysis of acoustic-like oscillations
of the field $p_e$.~\cite{arminjon93a} However, in this ether theory, an "absolute" version of
Einstein's equivalence principle occurs naturally.~\cite{arminjon93b} It leads to stating the
following {\em assumption (A):} When a gravitational field is present, {\em i.e.} in the case
of a non-uniform field $p_e$, any material object is contracted, only in the direction of the
field ${\sf g}$, in the ratio $\beta=\rho_e/\rho_e^\infty$, (where $\rho_e^\infty$ is the
value of $\rho_e$ "outside the gravitational field," {\em i.e.} in fact the upper bound of
$\rho_e$ over the whole space), and the period $T$ of any clock is dilated in the
corresponding ratio, {\em i.e.}, it becomes $T/\beta$. Due to this assumption, the Euclidean
"background" metric $g^0$ is different from the physical space metric $g$, which is a
Riemannian one, and the physical, local time $t_x$ differs from the "absolute time" $t$ in the
following way:
\begin{equation}
\frac{dt_x}{dt}=\beta(t,x),\; \drond{ }{t_x}=\frac{1}{\beta(t,x)}\! \drond{ }{t}.
\label{eq:temps-local}
\end{equation}
(Here $x$ is the spatial position in the preferred reference frame.) Moreover, SR leads to assuming the relation $p_e=c^2 \rho_e$. Accounting for this and for Eq.~(\ref{eq:temps-local}), the (preferred-frame) equation for the scalar gravitational field is finally stated in terms of the physical space metric and local time, thus
\begin{equation}
\Delta_g p_e - \frac{1}{c^2} \drond{^2 p_e }{t_x^2}= 4 \pi G \sigma \rho_e,
\label{eq:champ-scalaire}
\end{equation}
where $\sigma$ is the mass-energy density.~\cite{arminjon93b} More precisely, $\sigma$ is
defined as the $T^{00}$ component of the energy-momentum tensor of matter and nongravitational
fields {\bf T}, when the time coordinate is $x^0=ct$ with $t$ the absolute time, and in any
spatial coordinates that are adapted to the preferred reference frame.~\cite{arminjon96}
Motion of any test particle is governed by an extension of that form of Newton's second law
which is valid for SR. This extension involves the gravity acceleration vector ${\sf g}$, and
the velocity-dependent mass of SR, containing the Lorentz factor as a multiplying factor- the
velocity being measured with physical standards affected by gravitation as stated in
assumption (A) above.~\cite{arminjon96} The extension can be used for a dust continuum
(non-interacting matter), it leads then~\cite{arminjon98a} to the following (preferred-frame)
equation:
\begin{equation}
T_{\mu ;\nu}^\nu = b_\mu, \; b_0 \equiv \frac{1}{2}g_{jk,0}T^{jk}, \: b_i \equiv -
\frac{1}{2}g_{ik,0}T^{0k}. \label{eq:dynamique}
\end{equation}
(Semicolon means covariant derivative associated with the physical space-time metric $\gamma$; indices are raised or lowered using also metric $\gamma$.) Because gravitation is universal, Eq.~(\ref{eq:dynamique}) is assumed valid for any kind of continuous medium. Thus it allows, in particular, to define the gravitational modification of Maxwell's equations in the theory.~\cite{arminjon98b} This modification is consistent~\cite{arminjon98b} with photon dynamics as governed by the extended form of Newton's 2nd law. {\em Observational status of the theory:} the same gravitational effects on photons as in GR are predicted at the post-Newtonian approximation (PNA).~\cite{arminjon98c} An asymptotic PNA has been developed, allowing to build a consistent celestial mechanics in the theory.~\cite{arminjon00a,arminjon00b} The cosmic acceleration is predicted and nonsingular cosmological models are obtained.~\cite{arminjon99}

\eject
\end{document}